\documentclass[letterpaper,twocolumn,10pt]{article}
\usepackage{usenix}

\usepackage{graphicx}
\usepackage{fancyvrb}
\usepackage{listings}
\usepackage{listings-rust}
\usepackage{subcaption}

\newcommand{\name}{{\it Trust$<$T$>$}}
\begin{document}
\lstset{language=Rust, style=Boxed}
\title{\bf Delegation with Trust$<$T$>$: A Scalable, \\ Type- and Memory-Safe Alternative to Locks}
\date{}
\author{Noaman Ahmad <nahmad30@uic.edu>, Ben Baenen <bbaene2@uic.edu>, \\
Chen Chen <cchen262@uic.edu>, Jakob Eriksson <jakob@uic.edu>}
\maketitle
\begin{abstract}

  We present \name{}, a general, type- and memory-safe alternative to locking in concurrent programs. 
  Instead of synchronizing multi-threaded access to an object of type T with a lock, the programmer may place the object in a \name{}. The object is then no longer directly accessible. Instead a designated thread, the object's {\it trustee}, is responsible for applying any requested operations to the object, as requested via the \name{} API.
 
 Locking is often said to offer a limited throughput {\it per lock}.
 \name{} is based on delegation, a message-passing technique which does not suffer this per-lock limitation. 
 Instead, per-object throughput is limited by the capacity of the object's trustee, which is typically considerably higher.
 
 Our evaluation shows \name{} consistently and considerably outperforming locking where lock contention exists, 
 with up to 22$\times$ higher throughput in microbenchmarks, and 5--9 $\times$ for a home grown key-value store, as well as {\tt memcached}, in situations with high lock contention. Moreover, \name{} is competitive with locks even in the absence of lock contention. 
 \end{abstract}

\section{Introduction}
\label{s:intro}

Safe access to shared objects is fundamental to many multi-threaded programs.
Conventionally, this is achieved through {\it locking}, or in some cases through carefully designed lock-free data structures, both of which are implemented using atomic compare-and-swap (CAS) operations.
By their nature, atomic instructions do not {\it scale} well: atomic instructions must not be reordered with other instructions, often starving part of today's highly parallel CPU pipelines of work until the instruction has retired. This effect is exacerbated when multiple cores are accessing the same object, resulting in the combined effect of frequent cache misses and cores waiting for each other to release the cache line in question, while the atomic instructions prevent them from doing other work.

Delegation \cite{dice2011flath,calciu2013delegation, petrovic2015delegation,fatourou2011combining,hendler2010flat,oyama1999combining, yew1987combining, shalev2006combining, david2013everything,ffwd}, also known as message-passing or light-weight remote procedure calls (LRPC), offers a highly scalable alternative to locking.
Here, each shared object\footnote{Here, we use {\it object} to mean a data structure that would be protected by a single lock.} is placed in the care of a single core ({\it trustee} below).
Using a shared-memory message passing protocol, other cores (clients) issue requests to the trustee, specifying operations to be performed on the object.
Compared to locking, where threads typically contend for access, and may even suspend execution to wait for access, delegation requests from different clients are submitted to the trustee in parallel and without contention. This dramatically reduces the cost of coordination for congested objects.
The operations/critical sections are applied sequentially in both designs: by each thread using locks, or by the trustee using delegation; here delegation may benefit from improved locality at the trustee. 
Together, this translates to much higher maximum per-object throughput with delegation vs. locking.
Moreover, a single client thread may have multiple outstanding requests to one or more trustees, providing both parallelism and transparent batching benefits.

We propose \name{} (pronounced {\it trust-tee}), a programming abstraction and runtime system which provides safe, high-performance access to a shared object (or $property$) of type $T$.
Briefly, a \name{} provides a family of functions of the form: $$apply(c:FnOnce(\&mut~T)\rightarrow U)\rightarrow U,$$ which causes the closure $c$ to be safely applied to the property (of type $T$), and returns the return value (of type $U$) of the closure to the caller. 
Here, $FnOnce$ denotes a category of Rust closure types, and $\&mut$ denotes a mutable reference. 
(A matching set of non-blocking functions is also provided, which instead executes a callback closure with the return value.)

Critically, access to the property is only available through the \name{} API, which taken together with the Rust ownership model and borrow checker eliminates any potential for race conditions, given a correct implementation of {\it apply}. 
Our implementation of \name{} uses pure delegation. However, the design of the API also permits lock-based implementations, as well as hybrids. 

Beyond the API, \name{} provides a runtime for scheduling request transmission and processing, as well as lightweight user threads ($fibers$ below). This allows each OS thread to serve both as a Trustee, processing incoming requests, and a client. Multiple outstanding requests can be issued either by concurrent synchronous fibers or an asynchronous programming style.  

The primary contributions of this paper are as follows:

\begin{itemize}
  \setlength{\itemsep}{0em}
\item Trust$<$T$>$: a model for efficient, multi-threaded, delegation-based programming with shared objects leveraging the Rust type system.
\item A new delegation channel design, for delegating a variable number of arbitrary-sized and extremely flexible requests per message. 
\item Two efficient mechanisms for supporting nested delegation requests, a key missing ingredient in previous work on delegation.
\item Performance improvements up to 22$\times$ vs. the best locks on congested micro-benchmarks
\item Delegation performance consistently matching uncongested locks, given sufficient available parallelism
\item Memcached performance improvements of up to 9$\times$ on benchmarking workloads vs. stock {\tt memcached}.
  \end{itemize}

\section{Background and Motivation}
\label{s:background}

Locking suffers from a well-known scalability problem: as the number of contending cores grows, cores spend more and more of their time in contention, and less doing useful work. Consider a classical, but idealized lock, in which there are no efficiency losses due to contention. Here, the {\it sequential} cost of each critical section is the sum of (a) any wait for the lock to be released, (b) the cost of acquiring the lock, (c) executing the critical section, and (d) releasing the lock. Not counting any re-acquisitions on the same core, this must be at minimum one cache miss per critical section, in sequential cost. To make matters worse, this cache miss is incurred by an atomic instruction, effectively stallign the CPU until the cache miss is resolved (and in the case of a spinlock, until the lock is acquired).

Two main solutions to this problem exist. First, where the data structure permits, fine-grained locking can be used to split the data structure into multiple independently locked objects, thus increase parallelism, reduce lock contention and wait times. With the data structure split into sufficiently many objects, and {\it accesses distributed uniformly}, a fine-grained locking approach tends to offer the best available performance. 

The second solution is various forms of delegation, where one thread has custody of the object, and applies critical sections on behalf of other threads. Ideally, this minimizes the sequential cost of each critical section without changing the data structure: there are no sequential cache misses, ideally no atomic instructions, but of course the critical sections themselves still execute sequentially. 

{\it Combining} \cite{fatourou2012revisiting, hendler2010flat, dice2011flath, fatourou2011combining, oyama1999combining, yew1987combining, shalev2006combining}, is a flavor of delegation in which threads temporarily take on the role of {\it combiner}, performing queued up critical sections for other threads. Combining can scale better than locking in congested settings, but does not offer the full benefits of delegation as it makes heavy use of atomic operations, and moves data between cores as new threads take on the {\it combiner} role.  
Most recently, TCLocks \cite{tclocks} offers a fully transparent combining-based replacement for locks, by capturing and restoring register contents, and automatically pre-fetching parts of the stack. TClocks claims substantial benefits for extremely congested locks, and the backward compatibility is of course quite attractive. 
However, a cursory evaluation in \S\ref{s:eval} reveals that TCLocks substantially underperform regular locks beyond extremely high contention settings, and never approaches \name{} performance.

Beyond {\it combining}, delegation has primarily been explored in proof-of-concept or one-off form, with relatively immature programming abstractions. We propose \name{}, a full-fledged delegation API for the Rust language, which presents delegation in a type-safe and familiar form, while substantially outperforming the fastest prior work on delegation. 

While delegation offers much higher throughput for congested shared objects, it does suffer higher latency than locking in uncongested conditions.
To hide this latency, and make delegation competitive in uncongested settings, \name{} exposes additional concurrency to the application via asynchronous delegation requests and/or light-weight, delegation-aware user threads ({\it fibers}). 

Lacking modularity is another common criticism of delegation: in FFWD \cite{ffwd}, an early delegation design, delegated functions must not perform any blocking operations, which includes any further delegation calls. In \name{}, this constraint remains for the common case, as this typically offers the highest efficiency. However, \name{} offers several options for more modular operation. First, asynchronous/non-blocking delegation requests are not subject to this constraint - these requests may be safely issued in any context. Second, leveraging our light-weight user threads, we offer the option of supporting blocking calls in delegated functions, on an as-needed basis.

Finally, prior work on delegation has required one or more cores to be dedicated as delegation servers. While \name{} offers dedicated cores as one option, the \name{} runtime has every core act as a delegation server, again leveraging light-weight user threads. Beyond easing application development and improving load balancing, having a delegation server on every core allows us to implement \name{} without any use of atomic instructions, instead relying on delegation for all inter-thread communication. Beyond potential performance advantages, this also makes \name{} applicable to environments where atomic operations are unavailable. 

\section{\name{}: The Basics}
\label{s:design}
\begin{figure*}[ht]
  \begin{lstlisting}  
let ct = local_trustee().entrust( 17 );           // ct: Trust<i32>
ct.apply( |c| *c+=1 );                             //  c: &mut i32 
assert!(ct.apply( |c| *c ) == 18 );
\end{lstlisting}
  \caption{Minimal \name{} example. An entrusted counter, referenced by {\tt ct} is initialized to 17, then incremented once. 
  The comments on the right indicate the types of the variables.}
  \label{f:minimal}
\end{figure*}

The objective of \name{} is to provide an intuitive API for safe, efficient access to shared objects. 
Naturally, our design motivation is to support delegation, but the \name{} API can in principle also be implemented using locking, or a combination of locking and delegation. 
Below, we first introduce the basic \name{} programming model, as well as the key terms {\it trust, property, trustee} and {\it fiber} in the \name{} context, before digging deeper into the design of \name{}.

\subsection{Trust: a reference to an object}

A \name{} is a thread-safe reference counting smart-pointer, similar to Rust's {\it Arc$<$T$>$}. 
To create a \name{}, we clone an existing \name{} or $entrust$ a new object, or $property$ of type $T$, that is meant to be shared between threads. 
Once entrusted, the property can only be accessed by {\it applying} closures to it, using a trust.
Figure \ref{f:minimal} illustrates this through a minimal Rust example.
Line 1 entrusts an integer, initialized to 17, to the local trustee - the trustee fiber running on the current kernel thread. 
Line 2 applies an anonymous closure to the counter, via the trust.
The closure expected by {\tt apply} takes a mutable reference to the property as argument, allowing it unrestricted access to the property, in this case, our integer. 
The example closure increments the value of the integer. 
The assertion on line 3 is illustrative only. Here, we apply a second closure to retrieve the value of the
entrusted integer\footnote{A note on ownership: While the passed-in closure takes only a reference to the property, the Rust syntax $*c$ denotes an explicit dereference, essentially returning a copy of the property to the caller. This will pass compile-time type-checking only for types that implement $Copy$, such as integers.}.

\begin{figure*}[ht]
  \begin{subfigure}[t]{.5\textwidth}
    \begin{lstlisting}  
let ct = local_trustee().entrust(17);
let ct2 = ct.clone();
let thread = spawn(move || { 
 ct2.apply(|c| *c+=1);
});
ct.apply(|c| *c+=1);
thread.join()?;
assert!(ct.apply(|c| *c) == 19);
\end{lstlisting}   
\caption{Example using \name{}.}
    \label{f:minimal_threads}
  \end{subfigure}
  ~
  \begin{subfigure}[t]{.5\textwidth}
    \begin{lstlisting}  
let cm = Arc::new(Mutex::new(17));
let cm2 = cm.clone();
let thread = spawn(move || { 
 *(cm2.lock()?) += 1;
});
*(cm.lock()?) += 1;
thread.join()?;
assert!(*cm.lock()? == 19);
\end{lstlisting}
    \caption{The same program using standard Rust primitives. }
    \label{f:minimal_mutex}
  \end{subfigure}
  \caption{Minimal multi-threaded \name{} example. Reference counting ensures that the property remains in memory until the last \name{} referencing the property drops. }
\end{figure*}

In the example in Figure \ref{f:minimal_threads} the counter is instead incremented by two different threads.
Here, the {\tt clone()} call on {\tt ct} (line 2) clones the trust, but not the property; instead a reference count is incremented for the shared property, analogous to {\tt Arc::clone()}.
  On line 3, a newly spawned thread takes ownership of {\tt ct2}, in the Rust sense of the word, then uses this to apply a closure (line 4).
  When the thread exits, {\tt ct2} is dropped, decrementing the reference count, by means of a delegation request.
  When the last trust of a property is dropped, the property is dropped as well. 

For readers unfamiliar with Rust, Figure \ref{f:minimal_mutex} illustrates the rough equivalent of Figure \ref{f:minimal_threads}, but using conventional Rust primitives instead of \name{}. Note the similarity in terms of legibility and verbosity. 

\subsection{Trustee - a thread in charge of entrusted properties}

In our examples above, \name{} is implemented using delegation. Here, a {\it property} is {\it entrusted} to a {\it trustee}, a designated thread which executes applied closures on behalf of other threads. 
In the default \name{} runtime environment, every OS thread in use already has a trustee user-thread ({\it fiber}) that shares the thread with other fibers. When a fiber applies a closure to a trust, this is sent to the corresponding trustee as a message. Upon receipt, the trustee executes the closure on the property, and responds, including any closure return value. This may sound complex, yet the produced executable code 
substantially outperforms locking in congested settings.

A TrusteeReference API is also provided. Here, the most important function is {\tt entrust()}, which takes a property of type {\tt T} as argument (by value), and returns a \name{} referencing the property that is now owned by the trustee. This API allows the programmer to manually manage the allocation of properties to trustees, for performance tuning or other purposes. Alternatively, a basic thread pool is provided to manage distribution of fibers and variables across trustees.

\subsection{Fiber - a delegation-aware, light-weight user thread}

While the \name{} abstraction has some utility in isolation, it is most valuable when combined with an efficient message-passing implementation and a user-threading runtime. 
User-level threads, also known as coroutines or {\it fibers}, share a kernel thread, but each execute on their own stack, enabling a thread to do useful work for one fiber while another waits for a response from a trustee. This includes executing the local trustee fiber to service any incoming requests.

In this default setting, the synchronous {\tt apply()} function suspends the current fiber when it issues a request, scheduling the next fiber from the local ready queue to run instead. The local fiber scheduler will periodically poll for responses to outstanding requests, and resume suspended fibers as their blocking requests complete.

\subsection{Delegated context}
For the purpose of future discussion, we define the term {\it delegated context} to mean the context where a delegated closure executes. Generally speaking, closures execute as part of a trustee fiber, on the trustee's stack. 
Importantly, blocking delegation calls are not permitted from within delegated context, and will result in a runtime assertion failure. 
In \S\ref{s:nesting}, we describe multiple ways around this constraint.

\section{Core API}
\label{s:nesting}
The \name{} API supports a variety of ways to delegate work, some of which we elide due to space constraints. Below, we describe the core functions in detail. For a full API review, see the technical report and API documentation \cite{trust_techreport}. 

\subsection{{\tt apply()}: synchronous delegation}
\begin{lstlisting}[numbers=none]
apply(c: FnOnce(&mut T)->U)->U
\end{lstlisting}

{\tt apply()} is the primary function for blocking, synchronous delegaion as described in earlier sections. 
It takes a closure of the form {\tt |\&mut T| \{\}}, where {\tt T} is the type of the property. 
If the closure has a return value, apply returns this value to the caller. 

Importantly, {\tt apply()} is synchronous, suspending the current fiber until the operation has completed.
Often, the best performance with {\tt apply()} is achieved when running multiple application fibers per thread. 
Then, while one fiber is waiting for its response, another may productively use the CPU.

\subsection{{\tt apply\_then()}: non-blocking delegation}

\begin{lstlisting}[numbers=none]
  apply_then(c: FnOnce(&mut T)->U, 
          then: FnOnce(U))
\end{lstlisting}

\begin{figure*}
  \centering
    \begin{lstlisting}        
let ct = trustee.entrust(17);           // create trust for shared counter set to 17
ct.apply_then(|c| { *c+=1; *c },         // increment counter and return its value
              |val| assert!(val==18));  // check return value once received
\end{lstlisting}
\caption{Asynchronous version of the example in Fig. \ref{f:minimal}.
The second closure runs on the client, once the result of the first closure is received from the trustee. }
\label{f:minimal_async}
\end{figure*}

Frequently, asynchronous (or non-blocking) application logic can allow the programmer to express additional concurrency either without running multiple fibers, or in combination with multiple fibers. 
Here, {\tt apply\_then()} returns to the caller without blocking, and does not produce a return value.
Instead, the second closure, {\tt then}, is called with the return value from the delegated closure, once it has been received. 
Figure \ref{f:minimal_async} demonstrates the use of {\tt apply\_then()} following the pattern of Figure \ref{f:minimal}.

The {\tt then}-closure is a very powerful abstraction, as it too is able to capture variables from the local environment, 
allowing it to perform tasks like adding the return value (once available) to a vector accessible to the caller. 
Here, Rust's strict lifetime rules automatically catch otherwise easily introduced use-after-free and dangling pointer problems, forcing
the programmer to appropriately manage object lifetime either through scoping or reference counted heap storage. 

Importantly, as {\tt apply\_then()} does not suspend the caller, it may freely be called from within delegated context. 

\subsection{{\tt launch()}: apply in a trustee-side fiber}
\begin{lstlisting}[numbers=none]
  launch(c: FnOnce(&mut T)->U)->U
  launch_then(c: FnOnce(&mut T)->U, 
           then: FnOnce(U))

\end{lstlisting}

The most significant constraint imposed by \name{} on the closure passed to {\tt apply()} and {\tt apply\_then()} is that the closure itself may not block. 
Blocking in delegated context means putting the trustee itself to sleep, preventing it from serving other requests, potentially resulting in deadlock.
In previous work \cite{rcl}, this problem was addressed by maintaining multiple server OS threads, and automatically switching to the next server when one server thread blocks. This avoids blocking the trustee, but imposes high overhead, resulting in considerably lower performance, as demonstrated in \cite{ffwd}.

In \name{}, blocking in delegated context is prohibited: attempted suspensions in delegated context are detected at runtime, resulting in an assertion failure. Closures may still use {\tt apply\_then()}, but not the blocking {\tt apply()}. 
\footnote{Other forms of blocking, such as I/O waits or scheduler preemption, do not result in assertion failures. However, these can significantly impact performance if common, as blocking the trustee can prevent other threads from making progress.}

The lack of {\it nested blocking delegation} can be a significant constraint on the developer, and perhaps the most important limitation of \name{}. 
Specifically, it affects modularity, as a library function that blocks internally, even on delegation calls, cannot be used from within delegated context.  

\begin{figure*}
  \centering\
  \includegraphics[height=1.8in]{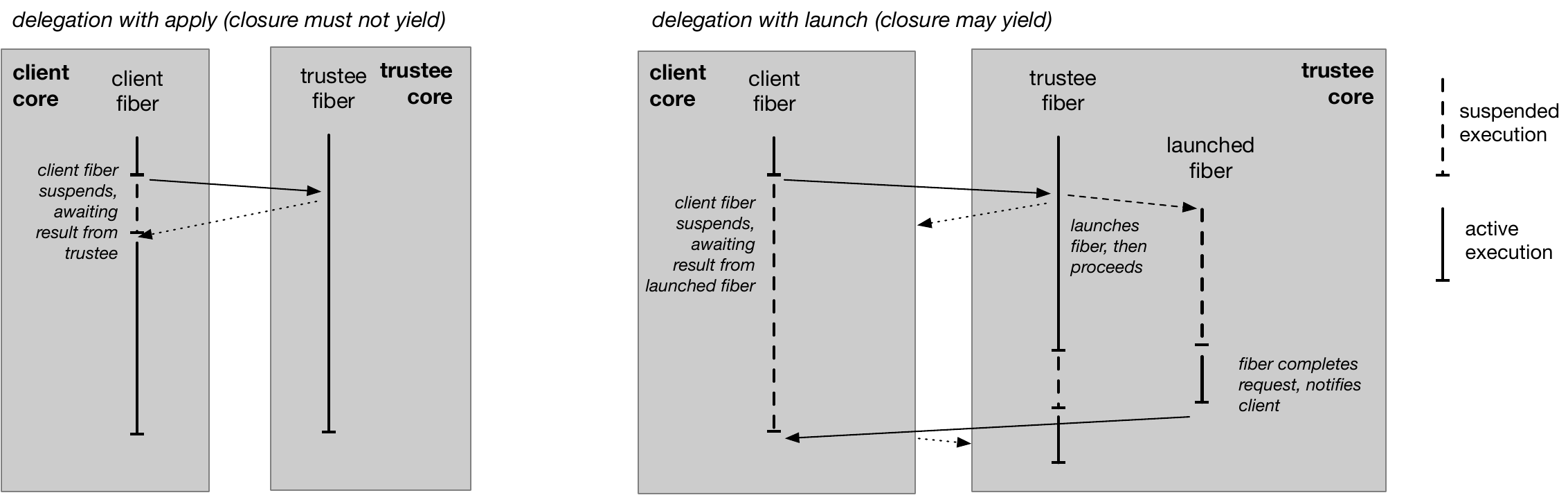}
  \caption{Operation of {\tt launch()} vs {\tt apply()}. {\tt launch()} supports blocking calls, including nested delegation calls in the
  delegated closure, but incurs a higher minimum overhead. Solid arrows indicate requests, dotted arrows are delegation responses.}
  \label{f:launch}
\end{figure*}

To address this, without sacrificing the performance of the more common case, we provide a convenience function: {\tt launch()}, which offers all the same functionality as {\tt apply()}, but without the blocking restriction.
Figure \ref{f:launch} describes {\tt launch()} from an implementation standpoint. {\tt launch()} creates a temporary fiber on the trustee's thread, which runs the closure. If this fiber is suspended, the client is notified, and the trustee continues to serve the next request. 
Once the temporary fiber resumes and completes execution of the closure, it then delivers the return value and resumes the client fiber via a second delegation call. Thus, if a delegated closure fails the runtime check for blocking calls, the developer can fix this by replacing the {\tt apply()} call, with a {\tt launch()} call. 


\subsubsection{Atomicity and {\tt launch()}}
That said, a complicating factor with blocking closures executed by {\tt launch()} is that without further protection, property accesses are no longer guaranteed to be atomic:
while the newly created fiber is suspended, another delegation request may be applied to the property, resulting in a race condition. 
To avoid this risk, {\tt launch()} is implemented only for {\tt Trust<Latch<T>{>}}.
{\tt Latch<T>} is a wrapper type which provides mutual exclusion, analogous to {\tt Mutex<T>} except that it uses no atomic instructions, and thus may only be accessed by the fibers of a single thread.\footnote{In Rust terms, {\tt Latch<T>} does not implement {\tt Sync}.}

\subsubsection{Leveraging Rust for safe and efficient delegation}
\label{s:noref}

Using the Rust type system, we ensure that delegated closures in \name{} cannot capture values that contain any references or pointers. 

In principle, this is far stricter than what is necessary: the existing and pervasive Rust traits {\tt Send} and {\tt Sync} already describe
the types that may be safely moved and shared between threads, and this continues to hold within \name{}. 

That said, safety does not guarantee performance. A common performance pitfall when writing delegation-based software is 
memory stalls on the trustee, which affects trustees disproportionately due to the polling nature of the delegation channel (see \S\ref{s:delegating}). Frequent cache misses and use of atomic instructions in delegated closures can substantially degrade trustee throughput vs. running closures with good memory locality. 

Generally speaking, cache line contention and use of atomic instructions are a natural result of sharing memory between threads. By prohibiting the capture of references and pointers, \name{} makes accidental shared memory patterns of programming much less likely in delegated code, and encourage pass-by-value practices. 

\subsubsection{Variable-size and other heap-allocated values}

Rust closures very efficiently and conveniently capture their environment, which {\tt apply()} sends whole-sale to the trustee. 
However, only types with a size known at compile time may be captured in a Rust closure (or even allocated on the stack). 

In conventional Rust code, variable size types, including strings, are stored on the heap, and referenced by a {\tt Box$<$T$>$} smart pointer. 
For the reasons described above (see \S\ref{s:noref}), we do not allow {\tt Box$<$T$>$} or other types that include pointers or references to be captured in a closure:
only pure values may pass through the delegation channel. 

As a result, variable size objects and other heap-allocated objects must be passed as explicit arguments rather than captured, so that they may be serialized before transmission over the delegation channel. 
For example, a {\tt Box$<$[u8]$>$} (a reference to a heap-allocated variable-sized array of bytes) cannot traverse the delegation channel. Instead, we encode a copy of the variable number of bytes in question into the channel, and pass this value to the closure when it is executed by the trustee. 
In practice, this takes the form of a slightly different function signature.

\begin{lstlisting}[numbers=none]
apply_with(c: FnOnce(&mut T, V)->U, w: V)->U
\end{lstlisting}

Here, the {\tt w:} argument is any type {\tt V:Serialize+Deserialize}, using the popular traits from the {\tt serde} crate. That is, any type that can be serialized and deserialized, may pass over the delegation channel in serialized form.
If more than one argument is needed, these may be passed as a tuple. Thus, to insert a variable-size key and value into an entrusted table, we might use:

\begin{lstlisting}[numbers=none]
  table_trust.apply_with(|table, (key, value)| 
    table.insert(key,value),(key,value))
\end{lstlisting}

We use the efficient {\tt bincode} crate internally for serialization.
As a result, while passing heap-allocated values does incur some additional syntax, the impact in terms of performance is minimal.



\section{Key Design and Implementation Details}

In this section, we delve deeper into the design and implementation of \name{}, from the mechanics of delegating Rust closures and handling requests and responses, to asynchronous versions of {\tt apply()}.

\subsection{Delegating Closures}
\label{s:delegating}

The key operation supported by \name{} is {\tt apply()}, which applies a Rust {\it closure} to the property referenced by the trust. A Rust closure consists of an anonymous function and a captured environment, which together is represented as a 128-bit {\it fat pointer}. Thus, to delegate a closure, a request must at minimum contain this fat pointer, and a reference to the property in question. 

One or more requests are written to the client's dedicated, fixed-sized {\it request slot} for the appropriate trustee.
That is, only the client thread may write to the request slot. 
For efficiency, if the captured environment of the closure fits in the request slot, we copy the environment directly to the slot, and update the fat pointer to reflect this change. 
A flag in the request slot indicates that new requests are ready to be processed. See \S\ref{s:slots} for details on request and response slot structure.

Responses are transmitted in a matching dedicated response slot. Leveraging the Rust type system, we restrict both requests and responses to types that can be serialized. 
The subtle implication of this is that the return value may not pass any references or pointers to trustee-managed data.\footnote{That said, we cannot prevent  determined Rust programmers from using {\tt unsafe} code to circumvent this restriction.}
While small closures with simple, known-and-fixed-size return types will generally yield the best performance, there is no limit beyond the serializability requirement on the size or complexity of closures and return types. 

\subsection{Scheduling Delegation Work}

Generally speaking, a call to {\tt apply()} appends a request to a pending request queue, local to the requesting thread.
In the case of {\tt apply()}, the calling fiber is then suspended, to be woken up when the response is ready.
Pending requests are sent during response polling, and as soon as an appropriate request slot is available. 
The intervening time is spent running other fibers, including the local trustee fiber, and polling for responses/transmitting requests.

There is a throughput/latency trade-off between running application fibers, and polling for requests/responses: poll too often, 
and few requests/responses will be ready, wasting polling effort. 
Poll too seldom, and many requests/responses will have been ready for a long time, increasing latency.
Automatically tuning this trade-off is an area of ongoing research. 
That said, the current implementation performs delegation tasks in a fiber that is scheduled in FIFO order just as other fibers.
After serving incoming requests, this fiber polls for incoming responses and issues any enqueued outgoing requests as applicable.

\subsubsection{Local Trustee Shortcut}

When a Trust has the current thread as its trustee, it is superfluous to use delegation to apply the closure. Instead, it is just as safe, and more efficient, to simply apply the closure directly, since we know that no other closures will run until the provided closure has run to completion. As a reminder, we know this because delegated closures may not suspend the current fiber.

\subsection{Request and Response Slot Structure}
\label{s:slots}

\begin{figure*}[t]
  \centering
  \includegraphics[width=5in]{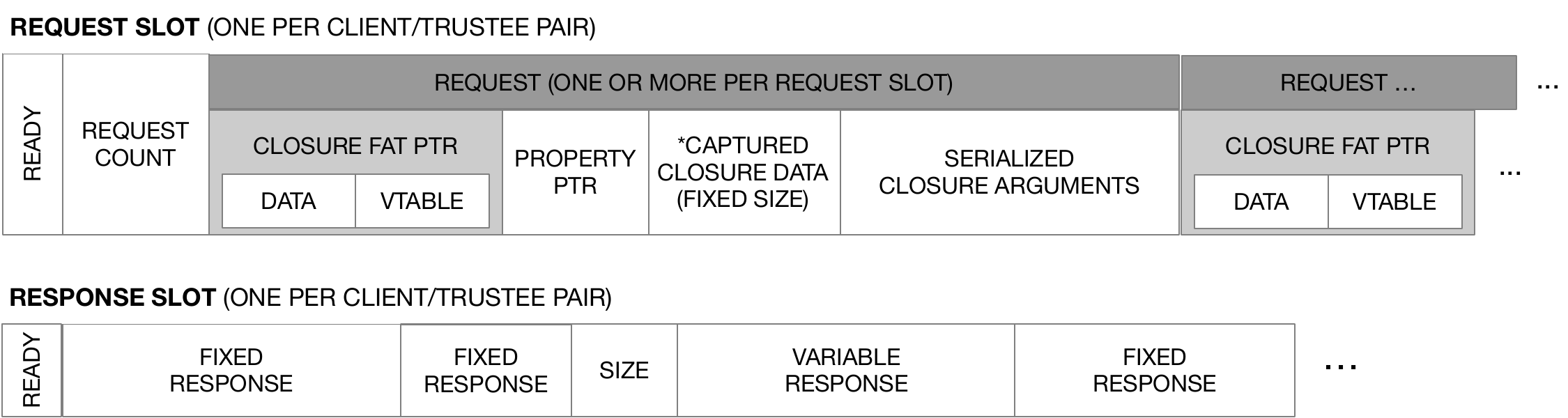}
  \caption{The fixed-size \name{} request slot consists of a {\tt ready} bit, a request counter, and a variable number of variable-sized requests. 
  The response slot contains a matching {\tt bit}, as well as one (fixed, variable, or zero-sized) response per request in the matching request slot. 
  There is one dedicated pair of request/response slots for each trustee/client pair.
  }
  \label{f:requestslot}
\end{figure*}
Figure \ref{f:requestslot} illustrates the internal structure of the basic request and response slot design. 
A header consisting of a {\tt ready} bit and a request count, is followed by a variable number of variable-sized requests.
The value of the {\tt ready} bit is used to indicate whether a new request or set of requests has been written to the slot: 
if the bit differs from the {\tt ready} bit in the corresponding response slot, then a new set of requests is ready to be processed.

By default, the slot size is 1152 bytes, and the client may submit as many closures as it can fit within the slot. Here, the minimum size of a request is 24 bytes: a 128-bit fat pointer for the closure, and a regular 64-bit pointer for the property.
The captured environment of Rust closures have a known, fixed size, which is found in the vtable of the closure. 
For typical small captured environments, this is copied into the request slot, and the pointer updated to point at the new location. 
Serialized closure arguments are appendend next, followed by the next request. 

Responses are handled in a similar fashion, though there is no minimum response size. 
Responses are sent simultaneously for all the requests in the request slot. 
The size of each response is often statically known, in which case it is not encoded in the channel.
Any variable-size responses are preceded by their size. 

The size of each request is always known, either statically or at the time of submission, which means we can restrict the number of requests sent to what can be accommodated by the request slot. 
The size of the response is not always known at the time the request is sent.
In cases where the combined size of return values exceeds the space in the response slot, the trustee dynamically allocates additional memory to fit the full set of responses,
at a small performance penalty. 

\subsubsection{Two-part slot optimization}

In order to accommodate a broad range of application characteristics, including those with a single trustee and many clients, as well as a single client with many trustees, we introduce a small optimization beyond the basic design above. 
Rather than represent the request and response slots as monolithic blocks of bytes, we represent each as two blocks: a 128-byte primary block, and a 1024-byte overflow block; each request and response is written, in its entirety, to one or the other block. 

This addresses an otherwise problematic trade-off with respect to the request and response slot sizes: 
with a monolithic request slot of, say, one kilobyte, the trustee would be periodically scanning flags 1024 bytes apart, a very poor choice from a cache utilization perspective, unless the slots are heavily utilized.
A two-part design accommodates a large number of requests (where needed), but improves the efficiency of less heavily utilized request slots by spacing ready flags, and a small number of compact requests, more closely using a smaller primary request block.

\section{Evaluation}
\label{s:eval} 

\begin{figure*}
  \begin{subfigure}[t]{.49\textwidth}
    \centering
\includegraphics[height=1.8in]{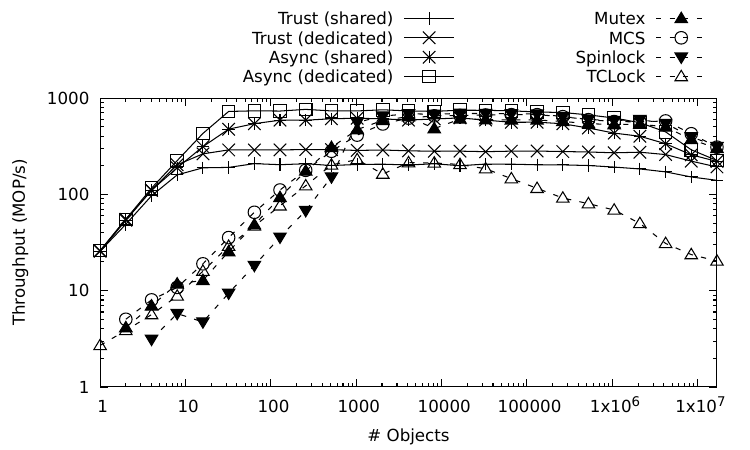}
\caption{Uniform access distribution. }
\label{f:fna_variables}
\end{subfigure}
\begin{subfigure}[t]{.49\textwidth}
  \centering
  \includegraphics[height=1.8in]{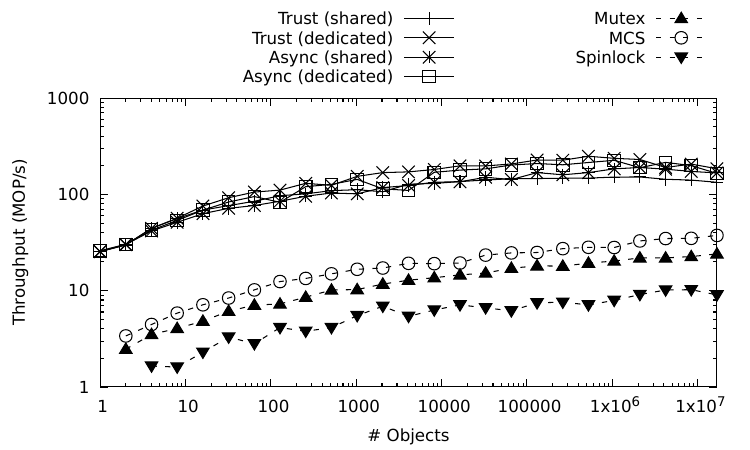}
  \caption{Zipfian access distribution, $\alpha=1$.}
  \label{f:fna_variables_zipf}
\end{subfigure}
\caption{Fetch-and-add throughput vs. object count. \name{} is substantially better than locks in congested settings,
and matches lock performance in uncongested settings. TCLocks were not found to be competitive.}
\end{figure*}

Below, we evaluate the performance of \name{} in two ways: 1) on both microbenchmarks, designed to stress test the core mechanisms behind \name{} and locking, and 2) on end-to-end application benchmarks,
which measure the performance impact of \name{} in the context of a complete system and a more realistic use case. 

\subsection{Fetch and Add: Throughput}

For our first microbenchmark, we use a basic fetch-and-add application. Here, a number of threads repeatedly increment a counter chosen from a set of one or more, and fetches the value of the counter. In common with prior work on synchronization and delegation \cite{dice2011flath,calciu2013delegation, petrovic2015delegation,fatourou2011combining,hendler2010flat,oyama1999combining, yew1987combining, shalev2006combining, david2013everything}, we also include a single {\tt pause} instruction in both the critical section and the delegated closures. The counter is chosen at random, either from a uniform distribution, or a zipfian distribution. Each thread completes 1 million such increments. 
In this section, each data point is the result of a single run.

Below, we primarily evaluate on a two-socket Intel Xeon CPU Max 9462, of the Sapphire Rapids architecture. This machine has a total of 64 cores, 128 hyperthreads, and 384 GB of RAM. Unless otherwise noted, we use 128 OS threads. In testing, several older x86-64 ISA processors have shown similar trends -- these results are not shown here.
 For locking solutions, we use standard Rust {\tt Mutex<T>} and the spinlock variant provided by the Rust {\tt spin-rs-0.9.8} crate, as well as MCSLock<T> provided by the Rust {\tt synctools-0.3.2} crate. For {\tt Trust<T>}, we show results for blocking delegation (Trust) as well as nonblocking delegation (Async).
In Fig. \ref{f:fna_variables}, we also include TCLocks, a recent combining approach offering a transparent replacement for standard locks, via the Litl lock wrapper \cite{litl} for {\tt pthread\_mutex}. 
To be able to evaluate this lock, we wrote a separate C microbenchmark, matching the Rust version. In the interest of an apples-to-apples comparison, we first verified that the reported performance with stock {\tt pthread\_mutex} on the C microbenchmark matched the Rust {\tt Mutex<T>} performance in our Rust microbenchmark.

Below, the {\tt Trust} results may be seen to represent any application with ample concurrency available in the form of conventional synchronous threads. {\tt Async} represents applications where a single thread may issue multiple simultaneously outstanding requests, e.g. a key-value store or web application server. 
Applications with limited concurrency are not well suited to delegation, except where the delegated work is itself substantial, which is not the case for this fetch-and-add benchmark. 
We further report results with both letting all cores serve as both clients and trustees ($shared$), as well as with an ideal number of cores dedicated serve only as trustees ($dedicated$).

\subsubsection{Uniform Access Pattern}

Figure \ref{f:fna_variables} illustrates the performance of several solutions on the uniform distribution version of this benchmark. 
For a very small number of objects, no data points are reported for some of the lock types - this is because the experiment took far too long to run due to severe congestion collapse. 

{\tt Trust<T>} substantially outperforms locking under congested conditions.
Between 1--16 objects, the performance advantage is 8--22$\times$ the best-performing MCSLock. 
For larger numbers of objects, the overhead of switching between fibers becomes apparent, as asynchronous delegation is able to reach a higher peak performance.  
In entirely uncongested settings, with 10$\times$ as many objects as there are threads, locking is able to match asynchronous delegation performance.
TCLocks \cite{tclocks} was the only lock type to complete the single-lock experiment within a reasonable time. 
It consistently outperforms spinlocks under congestion, and remains competitive with Mutex and MCS on highly congested locks.
However, TCLocks appear to trade their transparency for high memory and communication overhead, making it unable to compete performance-wise beyond highly congested settings. 
\footnote{TCLocks performance appears somewhat architecture dependent. In separate runs on our smaller Skylake machines, TCLocks were able to outperform Mutex by $\approx$50\% under the most extreme contention (a single lock).} 
Moreover, we struggled to apply TCLocks to memcached (which consistently crashed under high load), as well as to Rust programs (as Rust now uses built-in locks rather than {\tt libpthreads} wrappers). We thus elide TCLocks from the remainder of the evaluation.


\begin{figure*}[t]
  \begin{subfigure}[t]{.49\textwidth}
    \centering
  \includegraphics[height=1.8in]{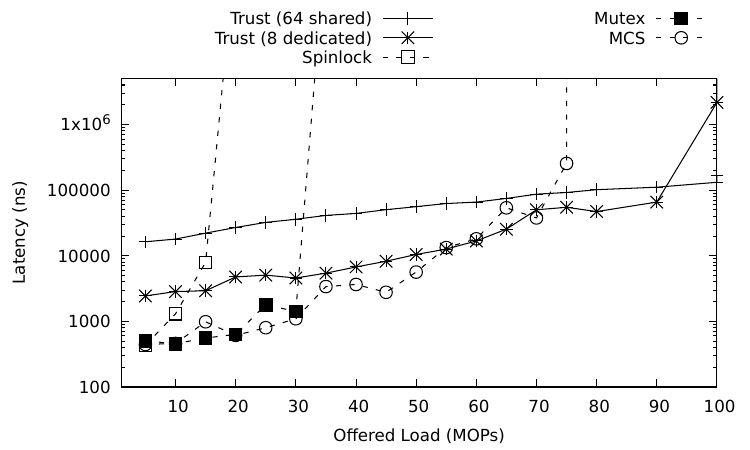}
  \caption{Uniform access distribution.}
  \label{f:latency_vs_load}
  \end{subfigure}
  \begin{subfigure}[t]{.49\textwidth}
    \centering
    \includegraphics[height=1.8in]{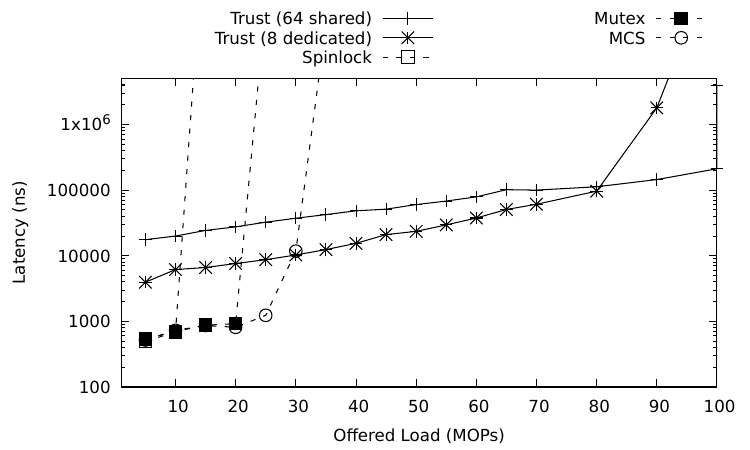}
    \caption{Zipfian access distribution, $\alpha=1$.}
    \label{f:latency_vs_load_zipf}
    \end{subfigure}
    \caption{Mean latency vs. offered load. In low-load settings, delegation incurs higher latency than locking. However, latency remains much more stable with increasing load. 
    The near-vertical lines show unbounded latency as capacity is reached.  
    }
  \end{figure*}

\subsubsection{Skewed Access Pattern: Zipfian distribution}

Zipf's law \cite{zipf1949} elegantly captures the distribution of words in written language. In brief, it says that the probability of word occurrence $p_w$ is distributed according to the rank $r_w$ of the word, thus: $p_w \propto {r_w}^{-\alpha}$, where $\alpha \sim 1$.
Similar relationships, often called ``power laws'', are common in areas beyond written language \cite{zipf1949, boltzmann, Ojovan_2006, DODDS20019, Venture}, sometimes with a greater value for $\alpha$. The higher the $\alpha$, the more pronounced is the effect of popular keys, resulting in congestion. 

Figure \ref{f:fna_variables_zipf} shows the results of our fetch-and-add experiment, but with objects selected according to a zipfian distribution ($\alpha=1$) instead of the uniform distribution above, representing a common skewed access distribution. 

With this skewed access pattern, \name{} overwhelmingly outperforms locking across the range of table sizes. 
This is explained by the relatively low throughput of a single lock. 
In our experiments, even MCSLocks, known for their scalability, offer at best 2.5 MOPs. 
When a skewed access pattern concentrates accesses to a smaller number of such locks, low performance is inevitable.
By comparison, a single \name{} trustee will reliably offer 25 MOPs, for similarly short critical sections.
For more highly skewed patterns, where $\alpha > 1$ (not shown), the curve grows ever closer to the horizontal as performance is bottlenecked by a small handful of popular items.

\subsection{Fetch and Add: Latency}

Next we measure mean latency for a scenario with 64 objects (uniform access distribution), and 1,000,000 objects (Zipfian access distribution), while varying the offered load. 
We show delegation results with 8 dedicated trustee cores, and with 64 shared trustee cores\footnote{The evaluation system has 64 cores, 128 hardware threads. In the vast majority of cases, having both hardware threads of each core work as trustees results in reduced performance.}.
We also plot the results for a spinlock, a standard Rust mutex, and an MCS lock as above. 

At low load, low contention results in low latency for locking, a ideal situation for locks.
However, as load increases, the locks eventually reach capacity, resulting in a rapid rise in latency. 
With \name{}, even low load incurs significant latency, due to message passing overhead. 
However, due to the much higher per-object capacity available, latency increases slowly with load until the capacity is reached. 
Thus, \name{} offers stable performance over a wide range of loads, at the cost of increased latency at low load. 
The higher latency does mean that to take full advantage of delegation, applications need to have 
ample parallelism available.


For both Uniform and Zipfian access distributions, we also measured 99.9th percentile (tail) latency (not shown). Overall, tail latency with locking (all types) tended to be approximately 10$\times$ the mean latency, in low-congestion settings. Delegation tail latency with a dedicated trustee, meanwhile, was 2.5$\times$ the mean, making delegation tail latency under low load only 2--3$\times$ that of locking.

It's also worth noting the difference between 8 dedicated trustees, and 64 trustees on threads shared with clients. The latency when sharing the thread with clients is naturally higher than when using trustees dedicated to trustee work. However, as load increases having more trustees available to share the load results in better performance. Using all the cores for trustees all the time also eliminates an important tuning knob in the system.




\subsection{Concurrent key-value store}
\label{s:kvstore}

\begin{figure*}[t]
  \centering
  \begin{subfigure}[t]{.45\textwidth}
      \centering
      \includegraphics[height=1.8in]{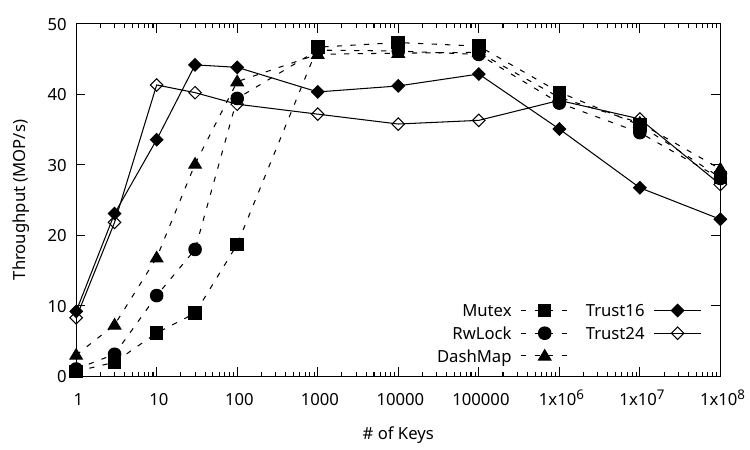}
      \caption{Uniform access distribution}
      \label{f:kvstore_tput}
  \end{subfigure}
  \begin{subfigure}[t]{.45\textwidth}
      \centering
      \includegraphics[height=1.8in]{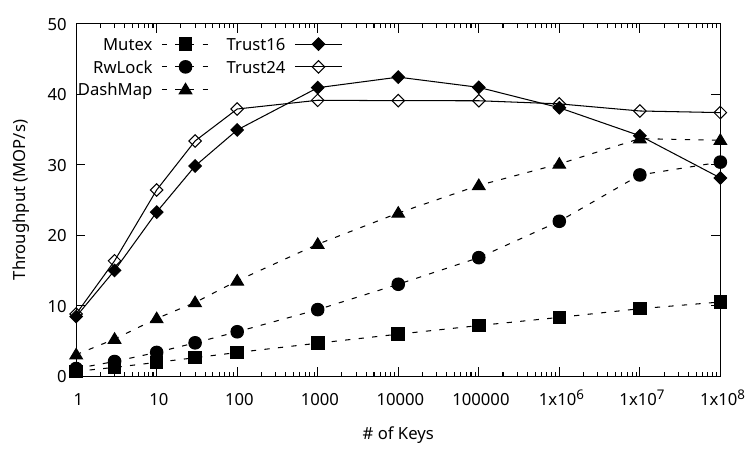}
      \caption{Zipfian access distribution}
      \label{f:kvstore_tput_zipf}
  \end{subfigure}

  \caption{Key-value store throughput, with 5\% writes and varying table size.}
\end{figure*}

\begin{figure*}[t]
  \centering
  \begin{subfigure}[t]{0.45\textwidth}
      \centering
      \includegraphics[height=1.8in]{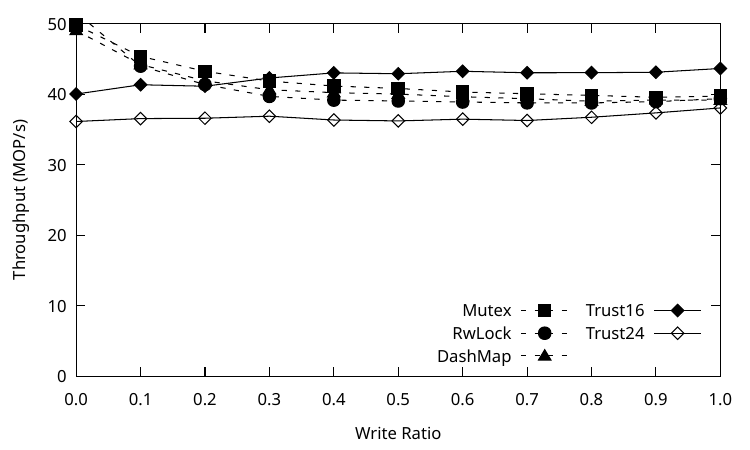}
      \caption{Uniform access distribution}
      \label{f:kvstore_tput_writes}
  \end{subfigure}
  \begin{subfigure}[t]{0.45\textwidth}
      \centering
      \includegraphics[height=1.8in]{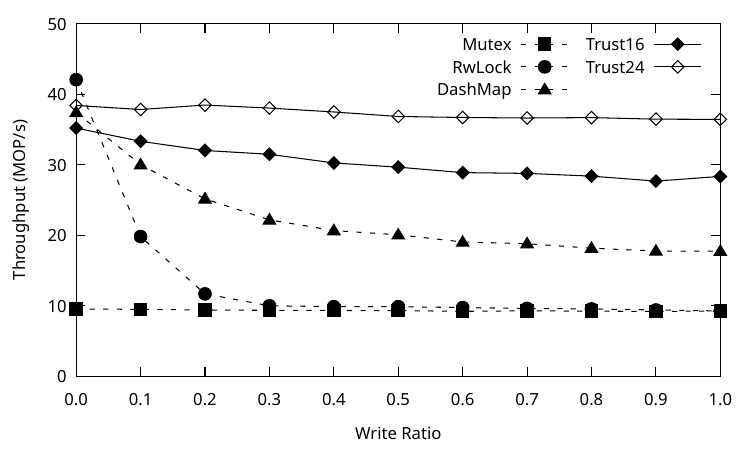}
      \caption{Zipfian access distribution}
      \label{f:kvstore_tput_writes_zipf}
  \end{subfigure}

  \caption{Key-value store throughput, with varying write percentage.}
\end{figure*}

For a more complete end-to-end evaluation, we implement a simple TCP-based key-value store, backed by a concurrent dictionary.
Here, we run a multi-threaded TCP client on one machine, and our key-value store TCP server on another, identical machine.
The two machines are connected by 100 Gbps Ethernet.
We compare our \name{} based solution to Dashmap \cite{dashmap}, one of the highest-performing concurrent hashmaps available as a public Rust crate,
as well as to our own na\"ively sharded Hashmap, using Mutex or Readers-writer locks and the Rust {\tt std::collections::HashMap$<$K, V$>$}.
Dashmap is a heavily optimized and
well-respected hash table implementation, which is regularly benchmarked against competing designs.

We implement the key-value store as a multi-threaded
server, where each worker-thread receives \texttt{GET} or \texttt{PUT} queries from one or more connections,
and applies these to the back-end hashmap. Both reading requests and sending results is done in batches, so as to
minimize system call overhead. Moreover, the client accepts responses out-of order, to minimize waiting.
The TCP client continuously maintains a queue of parallel queries over the socket, such that the server always has
new requests to serve.
In the experiments, we dedicate one CPU core to each worker thread.

For our sharded hashmaps, we create a fixed set of 512 shards, using many more locks than threads to reduce lock contention.
Dashmap uses sharding and readers-writer locks internally, but exposes a highly efficient concurrent hashmap API.
For our \name{} based key-value store, we use 16 and 24 cores to run trustees (each hosting a shard of the table) exclusively,
and the remaining cores for socket workers.
They are named Trust16 and Trust24, respectively.
Socket workers delegate all hash table accesses to trustees.
The key size is 8 bytes and the value size is 16 bytes in the experiments.
Prior to each run, we pre-fill the table, and report results from an average of 10 runs.

Figures \ref{f:kvstore_tput}--\ref{f:kvstore_tput_zipf} show the results from this small key-value store application,
for a varying total number of keys with 5\% write requests and 95\% read request,
and Uniform as well as Zipfian \cite{zipf1949} access distributions.
For Zipfian access, we use the conventional $\alpha=1$.
Overall, similar to the microbenchmark results, we find that the delegation-based solution
performs significantly better when contention for keys is high. However, due to the considerably higher complexity of
this application, the absolute numbers are lower than in our microbenchmarks. The relative advantage for delegation is also somewhat smaller,
as some parts of the work of a TCP-based key-value store are already naturally parallel.

For the Uniform distribution and 5\% writes, all the solutions perform similarly above 1,000 keys,
a large enough number that there is no significant contention. With 100 keys and less, \name{} enjoys a large advantage even under uniform access distribution.
With a Zipfian access distribution, accesses are concentrated to the higher-ranked keys, leading to congestion.
In this setting, \name{} trounces the competition, offering substantially higher performance across the full 1--100,000,000 key range.
It is interesting to note, also, that the Zipfian access distribution is where the carefully optimized design of
Dashmap shines, while it offers a fairly limited advantage over a na\"ive sharded design with readers-writer locks
on uniform access distributions. This speaks to the importance of efficient critical sections in the presence of lock congestion.

The througput of Trust16 is higher than Trust24 with 1,000--100,000 keys because it is of low cost to manage a relatively
small key space, while Trust16 can dedicate more resources to handle socket connections.
However, the performance of Trust16 starts to degrade with more keys, because the limited number of trustees fall
short when managing larger key spaces. With 24 trustees, the performance can be maintained at a high level.
The difference between Trust16 and Trust24 suggests an important direction of future research. For I/O heavy processes
like key-value stores, dedicated trustees will often outperform sharing the core between trustees and clients.
However, it is non trivial to correctly choose the number of trustees.
Automatically adjusting the number of cores dedicated to trustee work at runtime would be preferable.

In principle, readers-writer locks have a major advantage over \name{} in that they allow concurrent reader access,
while \name{} exclusively allows trustees to access the underlying data structure.
To better understand this dynamic, Figures \ref{f:kvstore_tput_writes}--\ref{f:kvstore_tput_writes_zipf}
show key-value store throughput over a varying percentage of writes.

Here, we use 1,000 keys for the Uniform access distribution, and 10,000,000 keys for Zipfian access distribution.
We note that these are table sizes where lock-based approaches hold an advantage in Figures \ref{f:kvstore_tput}--\ref{f:kvstore_tput_zipf}. 
For Uniform access patterns, where there is limited contention given the table size of 1,000 keys, the
impact of the write percentage is muted. For lock-based designs, the performance does drop somewhat, but remains
at a high level even with 100\% writes. 

It is interesting to note that \name{} performance increases modestly with the write percentage. One reason behind this is that in our 
key-value store, the closures issued by reads by necessity have large return values, while the closures issued by writes have no return values at all. This may allow the trustee to use only the first, small part of the return slot, occasionally saving two LLC cache misses per round-trip.


With the Zipfian access distribution, even with 10,000,000 keys, contention remains a bigger concern, especially for Mutex. All four designs exhibit reduced performance with increased write percentages, but again, \name{} 
proves more resilient. 
The efficiency advantage of Dashmap over our na\"ive lock-based designs is on full display with the Zipfian access 
distribution and a high write percentage. 
That said, the fundamental advantage of \name{} over locking in this application is clear.

\section{Legacy Application: Memcached}
\label{s:memcached}

We also port memcached version 1.6.20 to \name{} to demonstrate both the applicability and performance impact on legacy C applications.
Memcached is a multi-threaded key-value store application. Its primary purpose is serving PUT and GET requests with string keys and values over standard TCP/IP sockets. Internally, memcached contains a hash-table type data structure with external linkage and fine-grained per-item locking. By default, memcached is configured to use a fixed number of worker threads. Incoming connections are distributed among these worker threads. Each worker thread uses the {\tt epoll()} system call to listen for activity on all its assigned connection. Each connection to a memcached server traverses a fairly sophisticated state machine, a pipelined design that is aimed at maximizing performance when each thread serves many concurrent connections with diverse behaviors. The state machine will process requests in this sequence: receive available incoming bytes, parse one request, process the request, enqueue the result for transmission, and transmit one or more  results. 

For our port to \name{}, we eliminate the use of most locks, and instead divide the internal hash table and supporting data structures into one or more shards, and delegate each shard to one of potentially multiple trustees. 
Thus, instead of acquiring a lock, we delegate the critical section to the appropriate trustee for the requested operation.  
Our ported version follows the original state machine design, with one key difference:
for each incoming request on the socket, we make an asynchronous delegation request using {\tt apply\_then}, then 
move on to the next request without waiting for the response from the trustee.  
That is, rather than sequentially process each incoming request, we leverage asynchronous delegation to capture additional concurrency. 

A complicating factor in this asynchronous approach results from {\tt memcached} being initially designed for synchcronous operation with locking. 
For any one trustee-client pair, even asynchronous delegation requests are executed in-order, and responses arrive in-order.
However, this is not guaranteed for requests issued to different trustees. 
Consequently, the memcached socket worker thread must order the responses before they are transmitted over the network socket to the remote client.
By contrast, our delegation-native key-value store in \ref{s:kvstore} sends responses out of order over the socket, and instead includes a request ID in the response.

Another difference worth mentioning is that we don't allow delegation clients (in this case, the memcached socket worker thread) to access delegated data structures at all. 
This means that instead of a pointer to a value in the table, clients recieve a copy of the value. 
This significantly improves memory locality and simplifies memory management, since every value has a single owner. 
However, it does incur extra copying, which may reduce performance under some circumstances.\footnote{This can become a problem when values are large. For this use case, \name{} includes an equivalent of Rust's {\tt Arc<T>} which allows multiple ownership of read-only values. 
}

In practice, because memcached is written in C and \name{} is written in Rust, we cannot directly add delegation to the memcached source code. We address this in a two-step process: first, for any task that requires delegation, we create a 
minimal Rust function that performs that specific task. That is, a custom Rust function that becomes part of the 
memcached code base. Typically, such a function locates the appropriate Trust or TrusteeReference, and delegates a single closure.
Second, we break out the critical sections in the C code into separate inner functions that may be called from Rust. Thus, to delegate a C critical section, we simply call the inner function from a delegated Rust closure. 

Our port of Memcached to \name{} has approximately 600 lines of added, deleted or modified lines of code, out of 34,000+ lines total. 
This number includes approximately 200 of lines which were simply cut-and-pasted into the new inner functions for critical sections.
In addition, we introduced approximately 350 new lines of Rust code, to provide the interface between the C and Rust environments.

\subsection{Evaluation}
\begin{figure}[t]
  \centering
\includegraphics[height=1.8in]{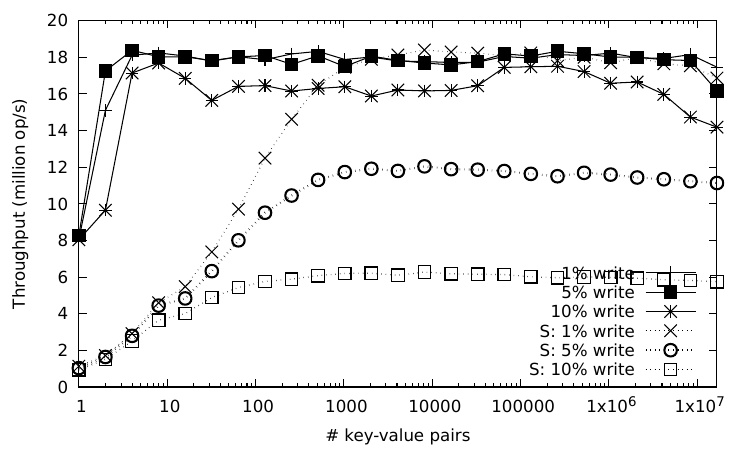}
  \caption{Memcached throughput with varying table size. Uniform access distribution. S: stock memcached. }
  \label{f:memcached_tput}
  \end{figure}

  \begin{figure}[t]
    \centering
    \includegraphics[height=1.8in]{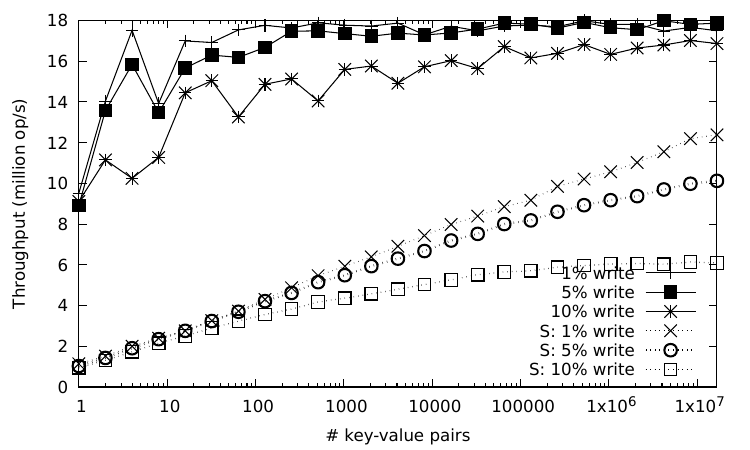}
    \caption{Memcached throughput with varying table size. Zipfian access distribution.}
  \label{f:memcached_tput_zipf}
  \end{figure}

To understand the performance of our delegated Memcached, we use the memtier benchmark client (version 1.4.0) with our delegated Memcached as well as stock memcached. 
For the cleanest results, but without loss of generality, we configure memcached with a sufficiently large hash power and available memory to eliminate table resizing and evictions. 
We also limit our evaluation to the conventional memcached PUT/GET operations.
Recent versions of memcached feature a optional new cache eviction scheme, which trades less synchronization for the need for a separate maintenance thread. For stock memcached, we evaluated both the traditional eviction scheme and the new one. We show results for the new scheme, which scales much better for write-heavy workloads and is otherwise similar in our setting. For our ported version, we use the traditional eviction scheme, maintaining one LRU per shard. 
Eviction is not relevant here, as we provide ample memory relative to the table size. 

The server and client run on separate machines, connected by 100Gbps Mellanox-5 Ethernet interfaces via a 100Gbps switch. Both client and server machines are 28-core, two-socket systems with Intel {\it Sandy Bridge} CPUs and 256 GB of RAM. The machines run Ubuntu Linux with kernel version 5.15.0.
Unless otherwise noted, we structure the experiments as follows: start a fresh {\tt memcached} instance. Populate the table with the indicated number of key-value pairs, then run measurements with 1\% writes, 5\% writes, and 10\% writes. 
After this, we start over with a new, empty {\tt memcached} instance. Each data point represents a single experiment, each set to last 20 seconds. For each, unless otherwise noted, we choose {\tt memcached} and {\tt memtier} parameters to maximize throughput. 
By default, this means 28 {\tt memcached} threads pinned to hardware threads 0--27. Running with 56 hardware threads did not yield any further performance improvement.
On the memtier side, we configure 28 threads, with four clients per thread, and pipelining set to 48.

Figures \ref{f:memcached_tput}--\ref{f:memcached_tput_zipf} illustrates the throughput of {\tt memcached} as we vary the number of keys in the table. 
While the absolute numbers are significantly lower than in the microbenchmarks and the key-value store, the overall picture from {\tt memcached} corresponds well with previous experiments. 

Using \name{} results in performance improvements of more than 5$\times$ when accessing popular objects, 
whether this popularity is due to a uniform access distribution across a smaller number of keys, or a Zipfian distribution over millions of of key-value pairs. 
When all items are accessed infrequently, locking suffers very little contention, and has the advantage of better distributing the work across cores. Here, this results in performance competitive with delegation, at least for read-heavy workloads.

The stock version is heavily affected by writes, due to the extra work required for these operations. This includes memory allocation, LRU updates as well as table writes, all of which involve synchronization in a lock-based design. With \name{}, all such operations are local to the shard/trustee, and do not require synchronization.  
With 5\% of writes, stock memcached loses $\approx$40\% of its performance, while the \name{} version sees only a minor performance penalty, resulting in delegation outperforming locking in this setting for the entire range of table sizes. While not shown, this trend continues with even more writes.

\section{Conclusions}

In this paper, we proposed Trust<T>, a new delegation-based programming model for safe, highly performance concurrent access to shared mutable state in Rust. Trust<T> provides an intuitive API that replaces locking with message passing between application threads and trustees in charge of shared data structures. 

Beyond the delegation API, we introduced two novel techniques to enable modular programming with nested delegation. First, apply\_then() are  non-blocking delegation requests, which may be issued from within a delegated context. Second, launch() safely supports arbitrary delegated code, by running the critical section in a separate fiber, protected by a single-threaded {\it latch} construct. 


Trust<T> provides evidence that delegation can be a competitive alternative to locking in real systems. The programming model integrates cleanly in Rust, making delegation an accessible option for developers. This work lays the groundwork for additional language and runtime support to unlock the performance and scalability benefits of delegation-based designs.
\label{s:conc}


\bibliographystyle{plain}
\bibliography{references}

\end{document}